\documentclass[aps,pre,reprint,superscriptaddress,amsmath,amssymb,showpacs,floatfix,longbibliography]{revtex4-1}
\usepackage{graphicx}
\usepackage{dcolumn}
\usepackage{bm}
\usepackage{color}
\usepackage[T1]{fontenc}
\usepackage[utf8]{inputenc}

\newcommand{\Tnet}{T_{L}}

\newcommand{\nup}{n_{+}}
\newcommand{\ndw}{n_{-}}
\newcommand{\nuu}{n_{++}}
\newcommand{\nud}{n_{+-}}
\newcommand{\ndd}{n_{--}}

\begin{document}
\title{Tricritical behavior of nonequilibrium Ising spins in fluctuating
environments}

\author{Jong-Min Park}
\affiliation{Department of Physics, University of Seoul, Seoul 02504, Korea}
\author{Jae Dong Noh}
\affiliation{Department of Physics, University of Seoul, Seoul 02504, Korea}
\affiliation{School of Physics, Korea Institute for Advanced Study,
Seoul 02455, Korea}

\date{\today}

\begin{abstract}
We investigate the phase transitions in a coupled system of Ising spins and
a fluctuating network. Each spin interacts with $q$ neighbors through links 
of the rewiring network. 
The Ising spins and the network are in thermal contact with the heat baths
at temperatures $T_S$ and $T_L$, respectively, so that the whole system is
driven out of equilibrium for $T_S \neq T_L$.
The model is a generalization of the $q$-neighbor Ising 
model~[A. J\k{e}drzejewski {\it et al.}, Phys. Rev. E {\bf 92}, 052105 (2015)], 
which corresponds to the limiting case of 
$T_L=\infty$. Despite the mean field nature of the interaction, 
the $q$-neighbor Ising model was shown to display a discontinuous phase
transition for $q\geq 4$. Setting up the rate equations for the
magnetization and the energy density, we obtain the phase diagram in the
$T_S$-$T_L$ parameter space. The phase diagram consists of 
a ferromagnetic phase and a paramagnetic phase. 
The two phases are separated by a continuous phase 
transition belonging to the mean field universality class or by a
discontinuous phase transition with an intervening coexistence phase. 
The equilibrium system with $T_S=T_L$ falls into the former case while the
$q$-neighbor Ising model falls into the latter case. At the tricritical
point, the system exhibits the mean field tricritical behavior. 
Our model demonstrates a possibility that a continuous phase transition
turns into a discontinuous transition by a nonequilibrium driving. 
Heat flow induced by the temperature difference between two heat baths is also
studied.
\end{abstract}

%64.60.Cn Order-disorder transitions
%05.70.Ln nonequilibrium Thermodynamics
%75.60.Nt Annealing magnetic
\maketitle

\section{Introduction}
The Ising model is one of the most studied statistical physics systems
for the theory of phase transitions and critical phenomena. 
Recently, J\k{e}drzejewski {\it et al}.~\cite{Jedrzejewski2015ga}
studied the phase transition in the so-called $q$-neighbor Ising model.
In this model, an Ising spin interacts ferromagnetically with 
$q$ instant neighbors which are chosen randomly among the other spins. 
The model was shown to undergo a phase transition
from a high-temperature paramagnetic phase to a low-temperature
ferromagnetic
phase. Interestingly, the phase transition is of first order~(discontinuous)
with a discontinuous jump in the spontaneous magnetization 
for $q\geq 4$, while it is of second order~(continuous) 
exceptionally at $q=3$. 

The $q$-neighbor Ising model looks similar to the Ising model on an 
annealed network~\cite{Lee2009ex}. 
Suppose that Ising spins are on nodes of a network 
and interact with each other through links.
In the annealed network, links are assumed to be rewired so fast that 
every spin is connected to all the others with effective coupling strengths. 
The equilibrium Ising model on the annealed network is described by the mean
field~(MF) theory and is shown to display the continuous phase
transition~\cite{Lee2009ex}.
In the $q$-neighbor Ising model, where spins interact with random neighbors, 
spatial correlations are negligible and the MF theory is also 
exact. 
Thus one might expect the continuous phase transition as the MF theory
predicts. Given the MF nature of the model, the discontinuous transition 
in the $q$-neighbor Ising model is puzzling.

The purpose of this study is to reveal the reason why the $q$-neighbor 
Ising model deviates from the equilibrium MF theory prediction.
We notice that not only the Ising spins but also the links connecting spins 
are fluctuating dynamic variables. The $q$-neighbor Ising model will be 
shown to be a limiting case of a nonequilibrium system driven between 
two heat baths $B_L$ and $B_S$ at different temperatures
$T_L$ and $T_S$, respectively. 
The Ising spins are in thermal contact with the heat bath $B_S$, while the
links are in thermal contact with $B_L$. The $q$-neighbor
Ising model corresponds to the case with $T_L=\infty$.
The nonequilibrium driving with $T_L\neq T_S$ is responsible for the
deviation from the equilibrium MF theory prediction. 

Phase transitions in nonequilibrium Ising models have been studied for 
a long
time~\cite{Masi1985ai,Miranda1987zf,Droz1990gc,Bassler1994ww,Szolnoki2000tb,Pleimling2010my,Borchers2014ci}. Ising spins can be driven out of equilibrium under any
dynamics breaking the detailed balance. The nature of resulting 
nonequilibrium phase transitions may or may not belong to the 
same universality class as the equilibrium counterpart.
The equilibrium Ising universality class is stable against a nonequilibrium
driving if the dynamics does not conserve the order 
parameter~\cite{Grinstein:1985ei,{Blote:1990ho}}. 
On the other hand, nonequilibrium Ising models with order parameter
conserving dynamics display different types of phase
transitions~\cite{{Schmittmann:1991da},{Schmittmann:1990ja},{Cheng:1991go},{Praestgaard:1994go},Bassler1994ww,Borchers2014ci}. 
Ising systems with spin-exchange dynamics are such examples. These systems
can be driven out of equilibrium by introducing multiple heat baths or 
a directional bias in the spin exchange process. 
In addition to the nonequilibrium critical phenomena, energy or particle 
currents~\cite{Borchers2014ci,Pleimling2010my} 
and the entropy production~\cite{Shim:2016dp} have been attracting growing interests recently.

The Ising spins in our study are 
connected via fluctuating links at a different temperature. 
In Sec.~\ref{model}, we introduce a nonequilibrium Ising model involving 
two heat baths of temperature $T_S$ and $T_L$. This model includes the 
$q$-neighbor Ising model as a limiting case.  
The analytic theory for the model is set up in Sec.~\ref{MFT},
and the resulting phase diagram in the parameter space of $T_S$ and $T_L$
is presented in Sec.~\ref{pd}. We find that the ordered phase and the
disordered phase are separated by the continuous phase transition line in
some region of the parameter space and by the coexistence phase in the other
region. The continuous phase transition line ends at the tricritical point.
The equilibrium model with $T_S = T_L$ undergoes the continuous phase
transition while the $q$-neighbor Ising model undergoes the discontinuous
phase transition through the coexistence phase.
We close the paper with summary and discussions on the heat flow 
in Sec.~\ref{summary}.

\section{Nonequilibrium Ising model}\label{model}
We begin with introducing the $q$-neighbor Ising model of
Ref.~\cite{Jedrzejewski2015ga}.
The system consists of $N$ Ising spins $s_n$~($n=1,2,\cdots,N$)
in thermal contact with a heat bath at temperature $T$. The spin states are
represented as $s_n = \pm 1$ or simply $\pm$. Spin configurations
are updated following the Monte Carlo rule. Each time step, one selects a
spin $s_i$ and $q$ other spins, denoted as $\{s_{i_k}|k=1,\ldots,q\}$, at
random. 
These $q$ spins are designated as instant interacting neighbors of 
$s_i$ with the energy function 
$E(s_i;\{s_{i_k}\}) = -J s_i \sum_{k=1}^q s_{i_k}$ 
with a ferromagnetic coupling constant $J>0$.
The spin $s_i$ is then flipped~($s_i\to -s_i$) with the probability
\begin{equation}\label{flip_prob}
P_T(\Delta E) \equiv \min\left[1,e^{-\Delta E/T}\right] \ ,
\end{equation}
where $\Delta E = E(-s_i;\{s_{i_k}\}) - E(s_i;\{s_{i_k}\})$ is
the energy change upon flipping $s_i$.

The flipping probability in \eqref{flip_prob} is taken commonly in
the Metropolis algorithm simulating the thermal equilibrium states 
at a given temperature $T/k_B$ with the Boltzmann constant
$k_B$~\cite{Binder:2010is}.
The Boltzmann constant will be set to unity hereafter.
Thus, the $q$-neighbor Ising model appears to be a thermal equilibrium 
system of Ising spins interacting with random neighbors. 
Surprisingly, the $q$-neighbor Ising model exhibits the
first-order phase transition for any $q\geq 4$~\cite{Jedrzejewski2015ga}.
The result is in sharp contrast to the equilibrium MF theory predicting 
the continuous phase transition~\cite{Goldenfeld:1992wv}.

In the $q$-neighbor Ising model, both the Ising spins and the
links between interacting spins are fluctuating dynamic variables. 
The Ising spins interact with the heat bath of temperature $T$. On the
other hand, the links are rewired completely randomly. This indicates
that two different heat baths, one for the spins and another for the
links, are involved in the $q$-neighbor Ising model.

To be more precise, we introduce the Hamiltonian for the whole
system including the spins and the links as
\begin{equation}\label{Hamiltonian}
H(\mathsf{A},\bm{s}) = - \frac{J}{2} \sum_{i,j} A_{i,j} s_i s_j \ ,
\end{equation}
where $A_{i,j}$ is an element of an adjacency matrix $\mathsf{A}$ and
$\bm{s} = (s_1,\ldots,s_N)$ denotes a spin configuration. The coupling
constant $J$ will be set to unity. 
The adjacency matrix element 
$A_{i,j}$ takes $1$ if there is a link between $i$ and $j$ 
and 0 otherwise. As a convention, we set $A_{i,i}=0$ disallowing a self-loop. 
The adjacency matrix is constrained by the condition 
\begin{equation}\label{constant_deg}
\sum_{j}A_{i,j} = q
\end{equation}
for all $i$ to ensure that every site has $q$ neighbors. 
Then, the $q$-neighbor Ising model is equivalent to the 
combined system of spins and links with the Hamiltonian 
\eqref{Hamiltonian} where 
the spins are in thermal contact with a heat bath ${B}_S$ 
of temperature $T_{S}=T$ and the links are in thermal contact with 
another heat bath ${B}_L$ of temperature $T_{L} = \infty$.

%%  Fig.   %%%%%%%%%%%%%%%%%%%%%%%%%%%%%%%%%%%%%%%%%%
\begin{figure}
\includegraphics*[width=0.9\columnwidth]{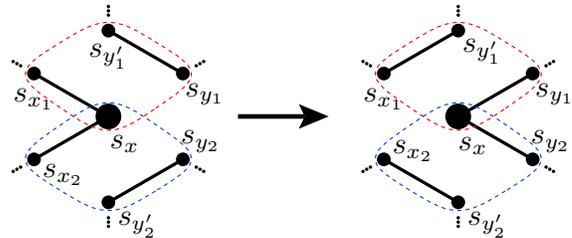}
\caption{Illustration of the link rewiring rule with $q=2$. 
Two quartets sharing $x$ are enclosed with dashed curves.
}
\label{fig1}
\end{figure}
%%%%%%%%%%%%%%%%%%%%%%%%%%%%%%%%%%%%%%%%%%%%%%%%%%%%%%

We now define the generalized model by introducing the following dynamics 
to the combined system with the Hamiltonian in~\eqref{Hamiltonian}.
The link configuration $\mathsf{A}$ and the spin configuration
$\bm{s}$ are updated as follows~(see Fig.~\ref{fig1}):
(i) Select a site $x$ at random. 
Current neighbors of $x$ are denoted as $x_1,\ldots,x_q$~($A_{x,x_k}=1$). 
One also selects $q$ distinct sites denoted as $y_1,\cdots,y_q$ 
among all sites but $x$. They are the potential candidates for new neighbors
of $x$. For each $y_k$, one further selects one of its neighbor $y'_k$ at
random~($A_{y_k,y'_k}=1$). 
(ii) Try to remove existing links between $x$ and $x_k$ and between $y_k$ and
$y'_k$~($A_{x,x_k}\to 0 \mbox{ and } A_{y_k,y'_k}\to 0$), and to add new links between $x$ and $y_{k}$ and between 
$x_k$ and $y'_k$~($A_{x,y_k}\to 1 \mbox{ and } A_{x_k,y'_k}\to 1$) 
for all $k=1,\ldots,q$. The link configuration after the
rewiring is denoted by $\mathsf{A}'$. 
The rewiring trial is accepted with
the probability $P_{T_L}(\Delta E)$, where $\Delta E =
H(\mathsf{A}',\bm{s})-H(\mathsf{A},\bm{s})$ is the energy change upon
rewiring with spin configuration $\bm{s}$ being fixed. 
(iii) The spin $s_x$ is then flipped to $-s_x$ with the probability
$P_{T_S}(\Delta E)$, where $\Delta E = H(\mathsf{A}'',\bm{s}_x) -
H(\mathsf{A}'',\bm{s})$ is the energy change upon spin flip. Here,
$\mathsf{A}''$ denotes the adjacency matrix after the rewiring
trial that is $\mathsf{A}'$ if the rewiring is accepted or 
$\mathsf{A}$ otherwise, and
$\bm{s}_x$ denotes the spin configuration with $s_x$ being flipped from
$\bm{s}$. The time $t$ is measured in unit of Monte Carlo step per site.

We adopt the so-called degree-preserving rewiring scheme in step
(ii)~\cite{{Maslov:2002hi},{Noh:2007cl}}.
This method allows one to rewire the links under the constraint of
\eqref{constant_deg}. Trials resulting in self-loops or 
double-links are rejected. 
When $T_L = \infty$, rewiring trials are always accepted.
Thus, our model with $T_S = T$ and $T_L=\infty$ reduces to the 
$q$-neighbor Ising model~\cite{Jedrzejewski2015ga}.
When $T_S=T_L$, the dynamics satisfies the detailed balance and the whole 
system is in thermal equilibrium~(see the discussion in Sec.~\ref{subs_db}).
When $T_L=0$, one may think that the Ising spins will be 
in thermal equilibrium on the quenched network. 
However, the network keeps evolving even at $T_L=0$. 
Suppose that the network reaches the ground state link configuration 
to a given spin configuration. 
When spin flips at finite $T_S$, the links are pumped out of
the ground state and rewired. Thus, the model with $T_L=0$ is different from 
the Ising model on the quenched network.

\section{Mean field theory}\label{MFT}
Link rewiring allows spins to interact with any other spins. 
Thus, spatial correlations between spins are negligible and the MF theory is
a good approximation. In this section, we derive the MF rate equations for the
mean magnetization density per site 
$m \equiv \frac{1}{N}\langle \sum_i s_i\rangle$ 
and the mean energy density per link
$e \equiv \frac{2}{qN} \langle H(\mathsf{A},\bm{s})\rangle$ taking account
of correlations up to nearest neighbors directly connected with links. 

We first introduce several notations.
Let $\nup$ and $\ndw$ be the fractions of $+$ and $-$ spins,
respectively.
The magnetization density is given by 
\begin{equation} 
m = \nup - \ndw \ .
\end{equation} 
The normalization $\nup+\ndw=1$ yields that 
\begin{equation}
n_{\pm} = \frac{1\pm m}{2} \ .
\end{equation}
Let $\nuu$, $\ndd$, and $\nud$ be the fractions of links
connecting $++$, $--$, and $+-$ spin pairs, respectively, satisfying the
normalization $\nuu+\ndd+\nud=1$.
The energy density per link is given by 
\begin{equation}
e = -(\nuu+\ndd)+\nud \ .
\end{equation}
Those fractions satisfy the relations 
$\nup = \nuu+\frac{1}{2}\nud$ and $\ndw = \ndd + \frac{1}{2}\nud$. 
Thus one can rewrite the fractions in terms of $m$ and $e$ as 
\begin{equation}\label{fraction}
\begin{split}
\nuu &= \frac{1}{4}{(1-e+2m)} \ , \\ 
\ndd &= \frac{1}{4}{(1-e-2m)} \ , \\
\nud &= \frac{1}{2}{(1+e)} \ . 
\end{split}
\end{equation}
Since $\nuu \geq 0$ and $\ndd \geq 0$, $e$ and $m$ are restricted within the
range $|m| \leq (1-e)/2$.

%%  Table.   %%%%%%%%%%%%%%%%%%%%%%%%%%%%%%%%%%%%%%%%%%
\begin{table}
\caption{\label{table1} Quartet configurations with $s_x=+$ 
and associated energy costs and realization probabilities. 
For $s_x=-$, $\omega^{-}_\alpha$ is the spin reversal of
$w^+_\alpha$ and $p^{-}_\alpha (m,e) = p^{+}_\alpha (-m,e)$.}
\begin{ruledtabular}
\begin{tabular}{cccccccc}
$\alpha$ & $\omega^{+}_\alpha$ &
$e_{\rm r}$ & $e_{\rm f}$ & $e_{\rm rf}$ &
$p^{+}_\alpha(m,e)$\\
\hline
\\[-1em]
1 & $(++++)$ &
0 & 2 & 2 &
$(1-e+2m)^2/16$\\
\\[-1em]
2 & $(+++-)$ &
0 & 2 & 2 &
$(1+e)(1-e+2m)/16$\\
\\[-1em]
3 & $(++-+)$ &
0 & 2 & -2 &
$(1+e)(1-e+2m)/16$\\
\\[-1em]
4 & $(++--)$ &
4 & 2 & -2 &
$(1-e+2m)(1-e-2m)/16$\\
\\[-1em]
5 & $(+-++)$ &
0 & -2 & 2 &
$(1+e)(1-e+2m)/16$\\
\\[-1em]
6 & $(+-+-)$ &
-4 & -2 & 2 &
$(1+e)^2/16$\\
\\[-1em]
7 & $(+--+)$ &
0 & -2 & -2 &
$(1+e)^2/16$\\
\\[-1em]
8 & $(+---)$ &
0 & -2 & -2 &
$(1+e)(1-e-2m)/16$
\end{tabular}
\end{ruledtabular}
\end{table}
%%  Table.   %%%%%%%%%%%%%%%%%%%%%%%%%%%%%%%%%%%%%%%%%%

Rewiring a single link of a randomly selected site $x$ 
involves a quartet of four spins $(s_x s_{x_k} s_{y_k} s_{y'_k})$~(see
Fig.~\ref{fig1}).
A quartet can take one of the $2^4=16$ spin configurations. 
We label the configurations
with $s_x=+1$ as $\omega^{+}_\alpha$~($\alpha=1,\ldots,8)$, and 
the spin-reversed configurations as $\omega^{-}_\alpha$. These
configurations are listed in Table~\ref{table1}. 
Given $m$ and $e$, the probability that a quartet is in a certain 
configuration $\omega^{\pm}_\alpha$ is given by a function of $m$ and $e$.
They will be denoted as $p^\pm_\alpha(m,e)$.
For example, the quartet $\omega^{-}_{8} = (-+++)$ has the probability 
$p^{-}_{8}= \frac{1}{2}\nud\nuu=\frac{1}{16}(1+e)(1-e+2m)$. 
The quartet probabilities are summarized in Table~\ref{table1}.
It is obvious that $p^+_\alpha(m,e)= p^-_\alpha(-m,e)$ and 
$\sum_{\alpha} p^{\pm}_\alpha(m,e) = n_{\pm} = (1\pm m)/2$.

It would cost
energy $e_{\rm r}(\alpha) =
\left [ -(s_x s_{y_k} + s_{x_k} s_{y'_k}) + (s_x s_{x_k} +
s_{y_k} s_{y'_k}) \right ]_{\omega^\pm_\alpha}$ for rewiring,
$e_{\rm f}(\alpha) = \left [ 2 s_x s_{x_k} \right ]_{\omega^\pm_\alpha}$ for
flipping of $s_x$ without rewiring, and $e_{\rm rf}(\alpha) =
\left [ 2 s_x s_{y_k} \right ]_{\omega^\pm_\alpha}$ for
flipping of $s_x$ after rewiring. The energy costs are summarized in
Table~\ref{table1}. Due to the spin-reversal symmetry of the Hamiltonian,
the energy costs for the configurations $\omega^+_\alpha$ and
$\omega^-_\alpha$ are the same.

Monte Carlo dynamics involves $q$ quartets sharing a randomly selected site
$x$. We denote the number of quartets of configuration $\omega^\pm_\alpha$ as
$n_\alpha$~$(=0,1,\ldots,q$). Due to the spin-reversal symmetry, we do not
need to count the number of quartets with $s_x=+1$ and $s_x=-1$ separately.
They are constrained by the sum rule $\sum_{\alpha=1}^8 n_\alpha = q$.

We are ready to set up the rate equations for $m$ and $e$. 
Suppose that a site $x$ is selected at random. 
Provided that $s_x=+$, the probability that $q$ associated quartets 
are specified by $\{n_\alpha\}$ is given by
\begin{equation}
f^+(m,e,\{n_\alpha\}) = \frac{q!}{(\nup)^q}
\prod_{\alpha=1}^8 \frac{(p^+_\alpha)^{n_\alpha}}{(n_\alpha)!} \ .
\end{equation}
The term $(\nup)^q=(\sum_\alpha p^{+}_\alpha)^q$ in the denominator 
guarantees the normalization $\sum_{\{n_\alpha\}}f^+(\{n_\alpha\})=1$,
where the summation $\sum_{\{n_\alpha\}}$ is over all sequences of
non-negative integers 
$\{n_\alpha\}$ satisfying $\sum_\alpha n_\alpha = q$.
Similarly, the probability for $\{n_\alpha\}$ with $s_x=-1$ is given by
\begin{equation}
f^-(m,e,\{n_\alpha\}) = \frac{q!}{(\ndw)^q}
\prod_{\alpha=1}^8 \frac{(p^-_\alpha)^{n_\alpha}}{(n_\alpha)!} \ .
\end{equation}
The symmetry property $p^{+}_\alpha(m,e)=p^{-}_\alpha(-m,e)$ yields that
\begin{equation}\label{f+-}
f^+(m,e,\{n_\alpha\}) = f^-(-m,e,\{n_\alpha\}) \ .
\end{equation}
The updating probabilities of links and spins are determined by the
associated energy changes.
The link rewiring would cost
\begin{equation}
E_{\rm r} = \sum_{\alpha} n_\alpha e_{\rm r}(\alpha) \ .
\end{equation}
The spin flip would cost 
\begin{equation}
E_{\rm f} = \sum_{\alpha} n_\alpha e_{\rm f}(\alpha) \ , 
\end{equation}
without link rewiring with probability $1-P_{T_L}(E_{\rm r})$, or
\begin{equation}
E_{\rm rf} = \sum_{\alpha} n_\alpha e_{\rm rf}(\alpha) \ ,
\end{equation}
after rewiring with probability $P_{T_L}(E_{\rm r})$.

Combining all the quantities, we finally obtain the 
rate equations in the $N\to\infty$ limit as
\begin{equation}\label{rate_eq}
\frac{dm}{dt} = F(m,e) \mbox{ and } \frac{de}{dt} = G(m,e) \ , 
\end{equation}
where
\begin{widetext}
\begin{equation}\label{FandG}
\begin{split}
F(m,e) &= 2 \sum_{\{n_\alpha\}} \left(- \nup f^+ + \ndw f^- \right)
   \left[P_{T_L}(E_{\rm r}) P_{T_S}(E_{\rm rf}) +
   \left\{1-P_{T_L}(E_{\rm r})\right\} P_{T_S}(E_{\rm f})
\right] \ , \\
G(m,e) &= \frac{2}{q}\sum_{\{n_\alpha\}} \left( \nup f^+ + \ndw f^- \right)
   \left[ E_{\rm r} P_{T_L}(E_{\rm r}) \left\{1-P_{T_S}(E_{\rm rf})\right\}
   + (E_{\rm r} + E_{\rm rf}) P_{T_L}(E_{\rm r}) P_{T_S}(E_{\rm rf}) 
     \right. \\
   & \left. \quad\quad \quad\quad\quad \quad\quad\quad \quad   \quad\quad \quad 
    + E_{\rm f} \left\{1-P_{T_L}(E_{\rm r})\right\} P_{T_S}(E_{\rm f})
\right] \ .
\end{split}
\end{equation}
\end{widetext}
Here, $P_T(E)$ is the
transition probability function defined in \eqref{flip_prob} and
the factor $(\frac{2}{q})$ of $G$ accounts for the link density.
The dependence on $T_L$, $T_S$, and $q$ is not shown explicitly.
Using the relation in \eqref{f+-}, one finds that 
\begin{equation}\label{FGsymm}
F(m,e) = -F(-m,e) \mbox{ and } G(m,e) = G(-m,e) \ .
\end{equation}
Note that in the $\Tnet \rightarrow \infty$ limit,
the function $F(m,e)$ becomes independent of $e$ and one recovers the
rate equation of Ref.~\cite{Jedrzejewski2015ga}.

\section{Phase diagram}\label{pd}
\begin{figure}
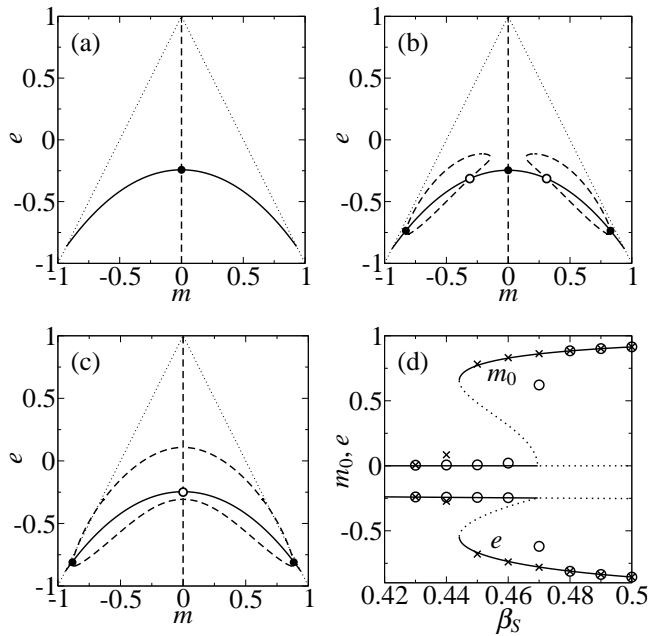

\includegraphics*[width=0.49\columnwidth]{fig2_a.eps}
\includegraphics*[width=0.49\columnwidth]{fig2_b.eps}
\includegraphics*[width=0.49\columnwidth]{fig2_c.eps}
\includegraphics*[width=0.49\columnwidth]{fig2_d.eps}
\caption{Fixed point analysis for $q=4$ and $\beta_L\equiv 1/T_L = 0.1$.
Nullclines with $F(m,e)=0$ (dashed line) and $G(m,e)=0$ (solid line) at
(a) $\beta_S \equiv 1/T_S = 0.44$, (b) $\beta_S=0.46$, and (c) $\beta_S=0.48$.
Dotted lines are the boundaries of the physical region $|m| \leq (1-e)/2$.
Closed and open circles indicate stable and unstable fixed points,
respectively.
(d) Spontaneous magnetization $m_0=|m|$ and the energy density $e$ 
at the stable~(solid line) and unstable~(dotted line) fixed points. 
Monte Carlo simulation data are also shown. The circle~(cross) symbols
represent the results in the cooling~(heating) setup with $N=10^6$ spins.}
\label{fig2}
\end{figure}

The steady-state phase diagram is determined by analyzing the fixed point
solution of the rate equation in \eqref{rate_eq}.
Firstly, Fig.~\ref{fig2} demonstrates how the fixed points bifurcate as $T_S$
varies with fixed $T_L=10$ at $q=4$. 
When $T_S>T_{d1}$ with a threshold temperature
$T_{d1}$, the system has a single stable fixed point at $m=0$~(see
Fig.~\ref{fig2}(a)) and is in a disordered paramagnetic phase. 
When $T_{d2} < T_S < T_{d1}$
with another threshold temperature $T_{d2}$,
two pairs of stable and unstable fixed points with $|m|\neq 0$
appear additionally~(see Fig.~\ref{fig2}(b)). 
Hence, the system can coexist in the paramagnetic 
phase and in the ordered ferromagnetic phase. 
When $T_{S} < T_{d2}$, the fixed point at $m=0$ becomes unstable after
merging with the unstable fixed points~(see Fig.~\ref{fig2}(c)). 
The system is in the ferromagnetic phase with nonzero spontaneous magnetization 
$m_0 \equiv |m|$. 
In Fig.~\ref{fig2}(d), we draw the steady state values of $e$ and $m_0$
against $T_S$. The system undergoes a first order transition
with the intermediate coexistence region.
This behavior is similar to that of the $q$-neighbor model
with $T_L=\infty$~\cite{Jedrzejewski2015ga}. 

The discontinuous transition is confirmed with the Monte Carlo
simulations. We have performed the simulations in two different setups.
In the cooling~(heating) setup, we increase~(decrease) the inverse 
temperature $\beta_S$ by $0.01$ in every 2000 time steps. The Monte Carlo
simulation data are presented in Fig.~\ref{fig2}(d). The numerical data
exhibit the hysteresis behavior which is characteristic of discontinuous
phase transitions.

\begin{figure}
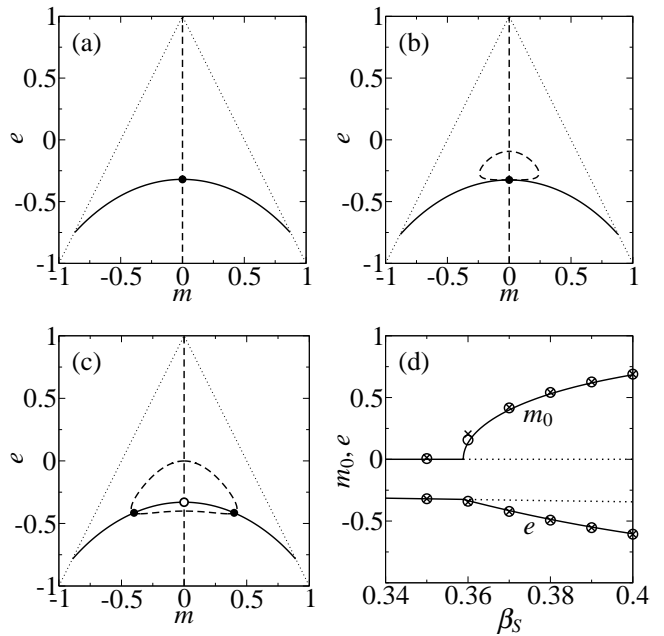

\includegraphics*[width=0.49\columnwidth]{fig3_a.eps}
\includegraphics*[width=0.49\columnwidth]{fig3_b.eps}
\includegraphics*[width=0.49\columnwidth]{fig3_c.eps}
\includegraphics*[width=0.49\columnwidth]{fig3_d.eps}
\caption{Fixed point analysis for $q=4$ and $\beta_L = 0.3$.
Nullclines at (a) $\beta_S = 0.35$,
(b) $\beta_S=1/T_c=0.3589$, and (c) $\beta_S=0.37$.
(d) Spontaneous magnetization $m_0$ and the energy density $e$ 
at the stable and unstable fixed points obtained from fixed point analysis.
The symbols represent the Monte Carlo simulation results. 
We use the same convention for lines and symbols as in Fig~\ref{fig2}.
}
\label{fig3}
\end{figure}

When we lower the temperature $T_L$, a qualitatively different behavior
emerges. Figure~\ref{fig3} shows the evolution of the fixed points as $T_S$
varies with fixed $T_L=10/3$. When $T_S > T_c$ with a critical threshold
temperature $T_c$, there is a single stable fixed point at $m=0$~(see
Fig.~\ref{fig3}(a)). As $T_S$ decreases below $T_c$, the fixed point at 
$m=0$ becomes unstable, while two stable fixed points with $m_0=|m|\neq 0$ 
appear near the unstable fixed point~(see Figs.~\ref{fig3}(b) and (c)). 
Hence, the spontaneous magnetization $m_0$ and the energy density $e$ vary
continuously and the system undergoes a continuous phase transition~(see
Fig.~\ref{fig3}(d)). 

The nature of the phase transition can be studied systematically. 
Let $\varepsilon(m)$ denote the nullcline satisfying 
$G(m,\varepsilon(m))=0$. The symmetry property $G(-m,e)=G(m,e)$ implies 
that the function is even in $m$, $\varepsilon(-m)=\varepsilon(m)$. 
The fixed points of the rate equation are found from zeroes of 
\begin{equation}
I(m) \equiv F(m,\varepsilon(m)) = -I(-m) \ .
\end{equation}
It is convenient to consider
\begin{equation}\label{Lfe}
L(m) = -\int_0^m I(m') dm' \ ,
\end{equation}
which is even in $m$. The stable fixed points of the rate equations
correspond to the local minima of $L(m)$. Hence, regarding $L(m)$ as the
Landau free energy, we can apply the phenomenological Landau
theory~\cite{Goldenfeld:1992wv}.
Note, however, that $L(m)$ is not the real free energy 
because the system is not in thermal equilibrium.

One can expand the Landau free energy as
\begin{equation}\label{Lexp}
L(m) = \frac{a_2}{2} m^2 + \frac{a_4}{4} m^4 + \frac{a_6}{6} m^6 +
O(m^8) 
\end{equation}
with $T_L$ and $T_S$ dependent coefficients $a_n$.
The paramagnetic fixed point at $m=0$ is stable when 
$a_2>0$ and unstable when $a_2<0$. 
Thus, the threshold for the paramagnetic state is
determined by the condition $a_2=0$. 
If $a_4$ is positive near the threshold, the spontaneous magnetization
scales as $m_0 \simeq (-a_2/a_4)^{1/2}$ and the system undergoes a continuous
phase transition. Figure~\ref{fig3} exemplifies this case.
On the other hand, if $a_4$ is negative near the threshold, the system is
bistable with $m_0=0$ and $m_0 \simeq [(-a_4+\sqrt{a_4^2-4 a_2
a_6})/(2a_6)]^{1/2}$ in the region $a_4^2 - 4 a_2 a_6 \gtrsim 0$.
The spontaneous magnetization jumps from zero to $m_0 \propto
(-a_4/a_6)^{1/2}$. Hence the system undergoes a discontinuous transition 
from the paramagnetic phase to the ferromagnetic phase separated by the
coexistence phase, as exemplified in Fig.~\ref{fig2}. The tricritical point
is located at the point where $a_2=a_4=0$.

\begin{figure}
\includegraphics*[width=\columnwidth]{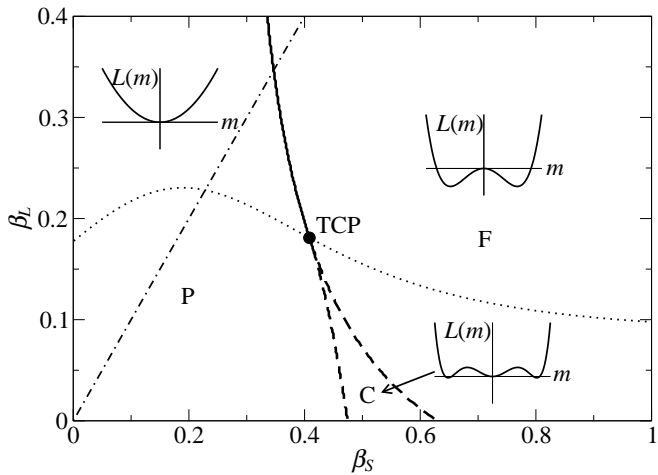}
\caption{Phase diagram at $q=4$. The solid line is the continuous phase
transition line between the P and the F phases. In the C phase, both
the ferromagnetic state and the paramagnetic state are stable. 
Along the dotted line, $a_4=0$. 
The dash-dotted line corresponds to the equilibrium line with 
$\beta_L=\beta_S$. 
Also shown is the shape of the Landau free energy $L(m)$ in each phase.
}
\label{fig4}
\end{figure}

We present the phase diagram for the system with $q=4$ in Fig.~\ref{fig4}.
The phase diagram consists of three phases: the paramagnetic~(P) 
phase, the ferromagnetic~(F) phase, and the coexistence~(C)
phase. The phase diagram is constructed as follows. We first draw the lines
$a_2=0$ and $a_4=0$. These lines are found numerically easily since we know
the analytic expressions for $I(m)$ and $L(m)$. The two lines intersect with 
each other at the tricritical point~(TCP). The line $a_2=0$ with $a_4>0$
is the boundary between the F and the P phases,
while the line $a_2=0$ with $a_4<0$
is the boundary between the F and the C phases.
The boundary between the P and the C phases,
which can be approximated by the line $a_4^2=4a_2a_6$ neglecting 
$O(m^8)$ term in \eqref{Lexp}, is located numerically by examining the
existence of the local minimum of $L(m)$ at $m\neq 0$.

\subsection{Equilibrium case with $T_S=T_L=T_{eq}$}\label{subs_db}
In order to reconcile with the results of the equilibrium Ising model 
on the annealed network~\cite{Lee2009ex}, we consider the equilibrium line
where $T_L=T_S=T_{eq}=1/\beta_{eq}$ in detail.
%%%%%%%%%%%%% revision
We can show that the transition probabilities in the rate equation satisfies
the detailed balance~(DB) condition. 

First, consider the rewiring process
which transforms each quartet configuration $\alpha=1,2,3,4,5,6,7,8$ to
$\gamma=\gamma(\alpha)= 1,2,5,6,3,4,7,8$, respectively. The DB
requires that
\begin{equation}
\frac{n_{\pm} f^{\pm}(m,e,\{n_\alpha\})}{n_{\pm}
f^{\pm}(m,e,\{n_{\gamma(\alpha)}\}} =
\frac{P_T(-\Delta E_r)}{P_T(\Delta E_r)} = e^{4\beta_{eq} (n_4 - n_6)} 
\end{equation}
with $\Delta E_{r} = 4(n_4-n_6)$~(see Table~\ref{table1}).
Using $p^{\pm}_3=p^{\pm}_5$ and $p^{\pm}_4/p^{\pm}_6 =
(1-e+2m)(1-e-2m)/(1+e)^2$, we find that the relation holds for all 
$\{n_\alpha\}$ if
\begin{equation}\label{db1} 
{(1-e+2m)(1-e-2m)} = e^{4\beta_{eq}}(1+e)^2 \ . 
\end{equation}

Secondly, consider the spin flip process which transforms each quartet 
configuration $\alpha=1,2,3,4,5,6,7,8$ to 
$\delta=\delta(\alpha)=8,7,6,5,4,3,2,1$, respectively.
The DB requires that
\begin{equation}
\frac{{n_\pm} f^{\pm}(m,e,\{n_\alpha\})}{ n_{\mp}
f^{\mp}(m,e,\{n_{\delta(\alpha)}\})} = \frac{P_T(-\Delta E_f)}{P_T(\Delta E_f)}
= e^{2\beta(n_a-n_b)} , 
\end{equation}
where $\Delta E_{f} = 2(n_a-n_b)$ 
with $n_a = (n_1+n_2+n_3+n_4)$ and $n_b = (n_5+n_6+n_7+n_8)$~(see
Table~\ref{table1}). Using the expressions for $p^{+}_\alpha(m,e) =
p^{-}_\alpha(-m,e)$ in Table~\ref{table1} and \eqref{db1}, 
we find that the relations holds for all $\{n_\alpha\}$ if
\begin{equation}\label{db2}
\left(\frac{(1-m)}{(1+m)}\right)^{2(q-1)} \left(
\frac{1-e+2m}{1-e-2m}\right)^q = 1  . 
\end{equation}

One can show further that the DB is also satisfied under the simultaneous
rewiring and flipping by combining the calculations for each. Therefore,
when $T_L=T_S$, the transition rates satisfy the DB condition and the
equilibrium energy density and the magnetization are determined by
\eqref{db1} and \eqref{db2}.

We add a remark on the DB. Although the DB condition is satisfied at the
rate equation level, it is not satisfied at the microscopic level of the 
Monte Carlo dynamics where the link rewiring and the spin flipping are tried
subsequently. 
One can show that the rewiring and flipping do not commute with each other,
which breaks the DB. Nevertheless, the preceding paragraphs show that 
the DB is satisfied in the average sense. Thus we will regard the model 
with $T_S=T_L$ as the equilibrium model.

\begin{figure}
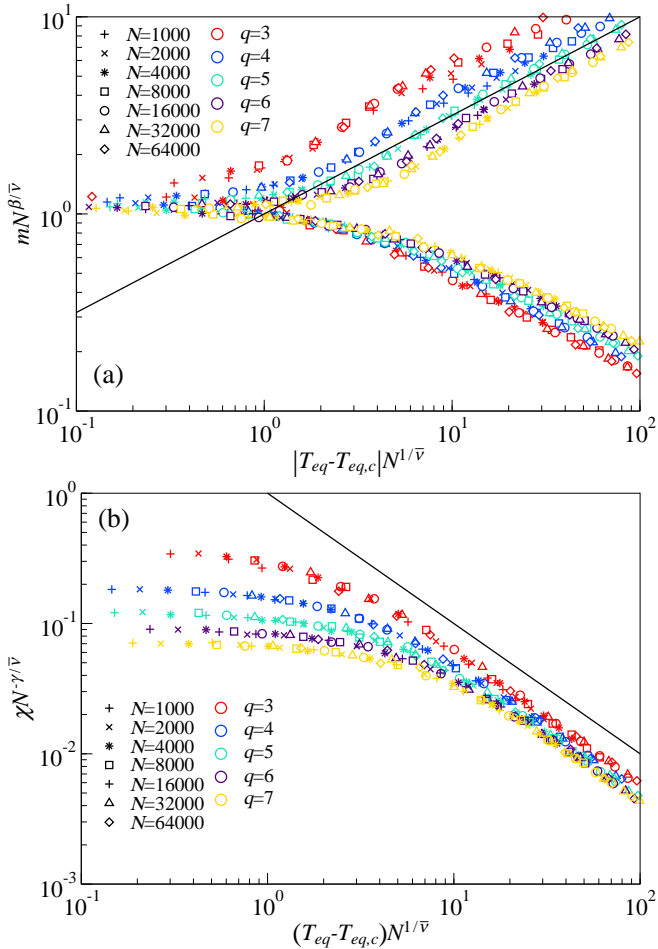

\includegraphics*[width=\columnwidth]{fig5_a.eps}
\includegraphics*[width=\columnwidth]{fig5_b.eps}
\caption{Finite size scaling analysis of the spontaneous magnetization 
$m_0$ in (a) and the susceptibility $\chi$ in (b) for the equilibrium model 
with $T_L=T_S=T_{eq}$ at $q=3$(red), $4$(blue), $5$(cyan), $6$(magenta), 
$7$(orange) according to \eqref{Teqc} and \eqref{scaling_func}. 
Each data set collapses well with the MF critical exponents $\beta = 1/2$, 
$\gamma = 1$, and $\bar{\nu} = 2$ near the critical point and for large
system sizes. 
The straight lines in (a) and (b) have the slopes $1/2$ and $-1$,
respectively.
}
\label{fig5}
\end{figure}

As seen from the phase diagram in
Fig.~\ref{fig4}, the equilibrium system undergoes the continuous phase
transition. The transition temperature $T_{eq,c}$ is found by analyzing
\eqref{db1} and \eqref{db2}. 
%Thus the critical temperature $T_{eq,c}$ is determined by 
%$a_2=0$ where $a_2 = \partial^2 L/\partial m^2 |_{m=0} = -\partial I/\partial
%m|_{m=0}$. This equation can be solved exactly for general $q$.
After a straightforward algebra, we obtain that
\begin{equation}\label{Teqc}
T_{eq,c} (q) = \frac{2}{\ln (q) - \ln (q-2)} \ .
\end{equation}
%The expansion in \eqref{Lexp} implies that 
The spontaneous magnetization 
behaves as $m_0 \sim |T_{eq}-T_{eq,c}|^\beta$
with the MF exponent $\beta = 1/2$.
We have also performed the Monte Carlo simulations to measure the other
critical exponents. Figure~\ref{fig5} shows the finite size scaling plots 
for the magnetization $m = \langle |\sum_i s_i | \rangle/N$ 
and the susceptibility
$\chi = ( \langle (\sum_i s_i)^2 \rangle
- \langle | \sum_i s_i | \rangle^2 )/(N T_{eq})$. Near the
critical point, they follow the scaling form 
\begin{equation}\label{scaling_func}
\begin{split}
m &= N^{-\beta/\bar{\nu}} O_m ( (T_{eq}-T_{eq,c}) N^{1/\bar{\nu}} ) \ ,  \\
\chi & = N^{\gamma/\bar{\nu}} O_\chi ( (T_{eq}-T_{eq,c}) N^{1/\bar{\nu}} ) 
\end{split}
\end{equation}
with scaling functions $O_m$ and $O_\chi$ and 
the MF critical exponents $\bar\nu = 2$ and $\gamma = 1$. The data collapse
confirms that the equilibrium model belongs to the MF universality class.
This result suggests that the discontinuous phase transition in the
$q$-neighbor model is the effect of the nonequilibrium driving.

%  Fig.   %%%%%%%%%%%%%%%%%%%%%%%%%%%%%%%%%%%%%%%%%%
%%%%%%%%%%%%%%%%%%%%%%%%%%%%%%%%%%%%%%%%%%%%%%%%%%%%%%

\subsection{Tricritical point}
The tricritical point TCP lies at the point where $a_2=a_4=0$ in \eqref{Lexp}.
The first condition $a_2=0$ yields that
\begin{equation}
I'(0)  = \left[(\partial_m + \varepsilon' \partial_e
)F\right]_{m=0,e=\varepsilon(0)} = 0 \ , 
\end{equation}
where $'$ denotes the derivative with respect to $m$ and 
$\partial_{m,e}$ is a shorthand notation for the partial differentiation.
Note that $\varepsilon(m)$ is an even function of $m$, hence
$\varepsilon'(0)=0$. Thus, we obtain the condition
\begin{equation}\label{TCP1}
\left[\partial_m F \right]_{m=0,e=\varepsilon(0)}=0 \ .
\end{equation}
The second condition $a_4=0$ requires that $I'''(0)=0$. Taking the
derivatives and using $\varepsilon'(0) = \varepsilon'''(0)=0$, one obtains
that $(\partial_m^3 + 3 \varepsilon''
\partial_m\partial_e)F(0,\varepsilon(0))=0$. The function $\varepsilon(m)$
is defined by the relation $G(m,\varepsilon(m))=0$, which yields that 
$\varepsilon''(0) =
-(\partial_m^2 G(0,\varepsilon(0)))/(\partial_e G(0,\varepsilon(0)))$. Thus,
we obtain that
\begin{equation}\label{TCP2}
\left[ (\partial_m^3 F) (\partial_e G) - 3 (\partial_m\partial_e F)
(\partial_m^2 G) \right]_{m=0,e=\varepsilon(0)}=0 \ .
\end{equation}
By solving \eqref{TCP1} and \eqref{TCP2}, we find the tricritical point is
located at
\begin{equation}
\beta_{L,{\rm TCP}} = 0.180954\cdots \mbox{ and } \beta_{S,{\rm TCP}}=
0.408461\cdots
\end{equation}
for $q=4$. The location of the TCP can be found numerically exactly for any
$q\geq 4$.  

\begin{figure}
\includegraphics*[width=\columnwidth]{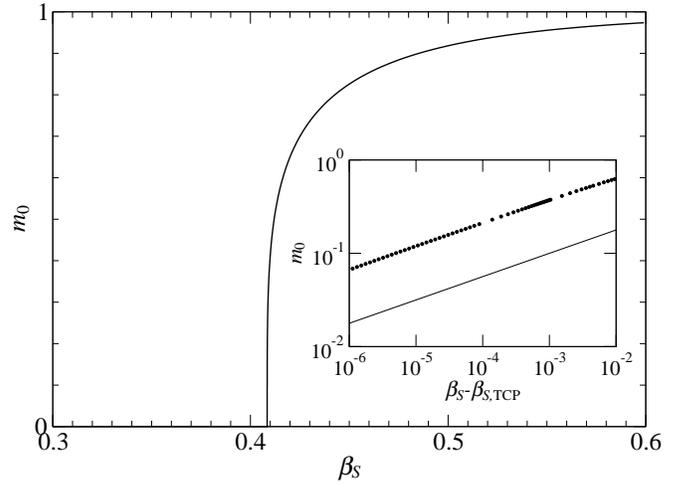}
\caption{Tricritical scaling of the spontaneous magnetization $m_0$ 
along the line $a_4=0$. 
The inset shows that $m_0\sim (\beta_S-\beta_{S,{\rm TCP}})^{1/4}$.}
\label{fig6}
\end{figure}
The order parameter $m_0$ follows the tricritical scaling behavior near 
the TCP. In Fig.~\ref{fig6}, we plot the spontaneous magnetization $m_0$
along the line $a_4=0$ shown in Fig.~\ref{fig4}. It scales as $m_0 \sim
(\beta_S-\beta_{S,{\rm TCP}})^{1/4}$ with the MF tricritical exponent $1/4$
instead of $1/2$~\cite{Goldenfeld:1992wv}.

At $q=3$, the lines $a_2=0$ and $a_4=0$ do not meet in the
$(\beta_S,\beta_L)$ space, and $a_4>0$ along the line $a_2=0$. Thus
the transition is always continuous and the tricritical point is absent. 

\section{Summary and discussions}\label{summary}
We have studied the phase transitions in the Ising spin system on 
the link-rewiring network. The system is in contact with two heat baths 
${B}_S$ and ${B}_L$ that govern the thermal fluctuations of the spins and the links,
respectively. This model is introduced in order to explain the
discontinuous phase transition recently reported in the $q$-neighbor 
Ising model where
Ising spins interact with random neighbors~\cite{Jedrzejewski2015ga}.
Such a result was puzzling since the MF theory is working in the 
$q$-neighbor Ising model and the equilibrium Ising model in the MF theory
exhibits a continuous phase transition. We have found that the
$q$-neighbor Ising model is indeed a nonequilibrium system driven between
two heat baths, ${B}_S$ for spins at finite temperatures and
${B}_L$ for links at the infinite temperature. 
We have constructed the phase diagram of the extended model 
in the parameter space of $\beta_S = 1/T_S$ and $\beta_L=1/T_L$ with the 
temperatures $T_S$ and $T_L$ for spins and links, respectively. 
When $T_S=T_L$, the model
reduces to the equilibrium model and displays the continuous phase
transition belonging to the equilibrium MF Ising universality class. 
When $T_L$ is much larger than $T_S$, the coexistence phase emerges and the
system exhibits the discontinuous phase transition. The coexistence phase
terminates at the tricritical point. Our result shows that the
nonequilibrium driving can change the nature of the phase transition from being
continuous to being discontinuous.

A thermal system in between two heat baths at different temperatures 
conducts heat from a high temperature bath to a low temperature one.
The steady-state heat flux, average heat flow per unit time, from the bath 
${B}_L$ to the system will be denoted as $\dot{Q}$. 
The steady-state heat flux from the bath 
${B}_S$ to the system is then equal to $-\dot{Q}$. 
The heat flow results in the increase of the total entropy 
with the rate $\dot{S} = \dot{Q} (-\beta_L + \beta_S)$. 
Recently, the
critical scaling behavior of the entropy production near the nonequilibrium
phase transition has been studied~\cite{Shim:2016dp}.
The heat is injected into the system from the bath ${B}_L$ 
when links are rewired. Hence, by
modifying \eqref{FandG}, one finds that the heat flux per link is written as 
\begin{equation}
\dot{Q} = \frac{2}{q}\sum_{\{n_\alpha\}} \left( \nup f^+ + \ndw f^- \right)
   E_{\rm r} P_{T_L}(E_{\rm r})  \ .
\end{equation}
The heat flux vanishes in the equilibrium case with $T_L=T_S$ due to the
detailed balance thereon.

%%%%%%%%%%%%%%%%%%%%%%%%%%%%%%%%%%%%%%%%%%%%%%%%%%%%%%%%%
\begin{figure}
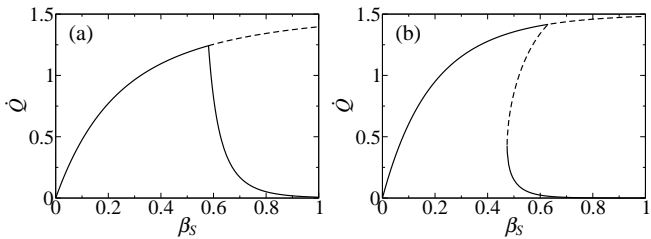

\includegraphics*[width=0.49\columnwidth]{fig7a.eps}
\includegraphics*[width=0.49\columnwidth]{fig7b.eps}
\caption{Heat flux from the heat bath ${B}_L$ at the infinite
temperature for $q=3$ (a) and $q=4$ (b). The solid lines represent the heat flux evaluated at
the stable fixed points, while the dashed lines represents the heat flux
evaluated at the unstable fixed points.
}
\label{fig7}
\end{figure}
%%%%%%%%%%%%%%%%%%%%%%%%%%%%%%%%%%%%%%%%%%%%%%%%%%%%%%%%%
We investigate the heat flow for the $q$-neighbor Ising model with 
$T_L=\infty$, where the expression is simplified to
\begin{equation}\label{dotQ}
\dot{Q} = -2q(e+m^2) \ .
\end{equation}
This expression is understood intuitively. 
Consider the rewiring of a single quartet. There are two links in a quartet, 
and the average energy of a quartet before rewiring is $2e$. 
After rewiring to random neighbors, the average energy becomes $-2m^2$.
Thus, the heat flux should be given by \eqref{dotQ}. 

The heat flux, evaluated from the fixed point solutions for $e$ and $m$, 
is presented in Fig.~\ref{fig7}. 
The nonzero positive heat flux confirms that the 
$q$-neighbor Ising model is indeed out of equilibrium. 
It varies continuously at $q=3$ and discontinuously at $q=4$ as the order
parameter $m_0$ does.
It is noteworthy that the heat flux in Fig.~\ref{fig7} 
increases~(decreases) as $\beta_S-\beta_L$ increase 
in the paramagnetic~(ferromagnetic) state. The heat flux usually
increases as the temperature difference becomes large. In this regard, the
decrease of $\dot{Q}$ in the ferromagnetic state is odd.
We also leave it for a future
work to understand the peculiar behavior.

\begin{acknowledgments}
This work was supported by the National Research Foundation of
Korea~(NRF) grant funded by the Korea government~(MSIP)
(No.~2016R1A2B2013972).
\end{acknowledgments}
 
\appendix
\bibliographystyle{apsrev}
\bibliography{paper}

\end{document}